\documentclass[10pt,conference]{IEEEtran}

\IEEEoverridecommandlockouts
\usepackage{amsmath,amssymb,amsfonts,mathtools}
\usepackage{algpseudocode}
\usepackage{algorithm}
\usepackage{graphicx}
\usepackage{textcomp}
\usepackage{xcolor}
\usepackage{mathptmx}
\usepackage{siunitx}
\usepackage{float}
\usepackage{epstopdf}
\usepackage{graphicx}
\usepackage{caption}
\usepackage{subcaption}
\usepackage{comment}
\usepackage{lipsum}
\usepackage{tikz,cite}
\usepackage{pgfplots}
\usetikzlibrary{decorations.pathreplacing,positioning}
\usepackage{caption,multirow,booktabs}
\usepackage{subcaption}
\usepackage[multiple]{footmisc}

\makeatletter
\def\ps@IEEEtitlepagestyle{%
  \def\@oddfoot{\mycopyrightnotice}%
}
\def\mycopyrightnotice{%
  \begin{minipage}{\textwidth}
  \centering \scriptsize
  \copyright 2025 IEEE. Personal use of this material is permitted. Permission from IEEE must be obtained for all other uses, in any current or future media, including reprinting/republishing this material for advertising or promotional purposes, creating new collective works, for resale or redistribution to servers or lists, or reuse of any copyrighted component of this work in other works.
  \end{minipage}
}
\makeatother

\begin{document}

\title{Spatially Consistent Air-to-Ground Channel Modeling with Probabilistic LOS/NLOS Segmentation

\thanks{Aymen Fakhreddine's contribution is funded by the Austrian Science Fund (FWF\,--\,Der Wissenschaftsfonds) under grant ESPRIT-54 (Grant DOI: 10.55776/ESP54).}
}

\author{\IEEEauthorblockN{Evgenii Vinogradov$^{1,2}$, Abdul Saboor$^{2}$, Zhuangzhuang Cui$^{2}$, Aymen Fakhreddine$^{3}$}
\IEEEauthorblockA{$^1$\textit{NaNoNetworking Center in Catalonia (N3Cat), Universitat Polit\`{e}cnica de Catalunya, Spain};\\ $^2$\textit{Department of Electrical Engineering, KU Leuven, Belgium};\\ $^3$\textit{Institute of Networked and Embedded Systems, University of Klagenfurt, Austria} \\
Email: evgenii.vinogradov@upc.edu}
}

\maketitle

\begin{abstract}
In this paper, we present a spatially consistent A2G channel model based on probabilistic LOS/NLOS segmentation to parameterize the deterministic path loss and stochastic shadow fading model. Motivated by the limitations of existing Unmanned Aerial Vehicle (UAV) channel models that overlook spatial correlation, our approach reproduces LOS/NLOS transitions along ground user trajectories in urban environments. This model captures environment-specific obstructions by means of azimuth and elevation-dependent LOS probabilities without requiring a full detailed 3D representation of the surroundings. We validate our framework against a geometry-based simulator by evaluating it across various urban settings. The results demonstrate its accuracy and computational efficiency, enabling further realistic derivations of path loss and shadow fading models and thorough outage analysis.
\end{abstract}


\section{Introduction}

Providing global seamless connectivity is a key objective of 6G~\cite{Dang2020_6G}. Non-Terrestrial Networks (NTNs) are considered vital enablers for supporting diverse mobility services by integrating satellite and aerial platforms together with terrestrial infrastructure~\cite{Geraci2022ntn}. Our previous work~\cite{Saboor2024its,Saboor2025pedestrian} both stressed the importance of Unmanned Aerial Vehicles (UAVs) in enhancing the connectivity of both pedestrian and vehicular users in the context of Intelligent Transportation Systems (ITS).

The Air-to-Ground (A2G) communication link performance is determined by the Line-of-Sight (LOS) availability. LOS presence critically impact signal strength and overall system reliability. Existing models for UAV channels often focus on evaluating the LOS probability based on simplified urban layouts~\cite{Saboor2023simulator,Saboor2024model}. While these approaches provide useful insights, such models neglect the spatial aspect, which accounts for the preservation of correlated LOS/NLOS conditions along the user's trajectory. The authors of~\cite{Karttunen2017spatial} demonstrate the importance of spatial consistency for the accuracy of terrestrial LOS and channel models, particularly at high frequencies.

Multiple works have surveyed UAV channel modeling from various perspectives. For instance,~\cite{khuwaja2018survey} illustrates the need for realistic channel models that account for mobility, antenna orientation, and environment-specific characteristics. It particularly emphasizes the importance of capturing spatial and time-varying behaviors in channel models, especially for mobile users in urban environments. The recent work~\cite{giuliani2024spatially} introduced a framework based on Generative Adversarial Network (GAN) for spatially consistent channel modeling. This model depicts the large-scale channel behavior as the local-average received signal strength (RSS) of a UAV navigating typical urban trajectories. The similarity to our work consists in relying on Ray Tracing (RT) data, however, it focuses on the scenario where the UAV is a User Equipment connected to ground base stations, which is fundamentally different from the scenario under study in our work.

In this work, we introduce an ITU-inspired~\cite{ITU} probabilistic approach to reproduce the transitions between NLOS and LOS conditions. We validate the spatial LOS behavior with our lightweight RT simulator introduced in~\cite{Saboor2023simulator}. Next, the modeled LOS states are used to reproduce the spatially consistent behavior of the A2G wireless channel.

\section{System model}\label{sec:system}
A single aerial base station (ABS) is deployed above a built-up urban area to provide communication services to a ground user equipment (UE). The ABS is located at horizontal coordinates $\mathbf{x}^{ABS} = (x^{ABS}, y^{ABS})$ and hovers at height $h^{ABS}$. The UE moves along a ground-level path $\mathbf{P}$, consisting of discrete points $\mathbf{p}_{i} = (x^{UE}, y^{UE})$. We focus on spatial modeling; time and velocity are not explicitly modeled, as results can be generalized to dynamic UEs via distance-to-time mappings.

\subsection{Channel model}
Communication occurs between the ABS and a ground-based UE as it moves along the path. At any UE location $\mathbf{p}_i$, the {attenuation caused by the large-scale channel components} (in dB) is given by:
\begin{equation}\label{eq:total_channel}
\Lambda(\mathbf{p}_{i}) = \Lambda_0 + \Lambda_{\text{ex}}(\mathbf{p}_{i}) + \xi(\mathbf{p}_{i}),
\end{equation}
where $\Lambda_0$ is the free-space path loss (FSPL) at a reference distance $d_0$ and frequency $f$, $\Lambda_{\text{ex}}(\mathbf{p}_i)$ is the excess path loss that depends on the link geometry, including the presence or absence of a line-of-sight (LOS), and $\xi(\mathbf{p}_i)$ is a zero-mean shadow fading term. The shadow fading is spatially correlated with an autocorrelation function:
\begin{equation}\label{eq:acf}
    R(\Delta d) = e^{-\frac{\Delta d}{d_{decorr}}},
\end{equation}
where $\Delta d$ is the spatial distance between two points and $d_{decorr}$ is the decorrelation distance.

\subsection{Communication link geometry}

\begin{figure}
    \centering
    \includegraphics[width=0.8\columnwidth]{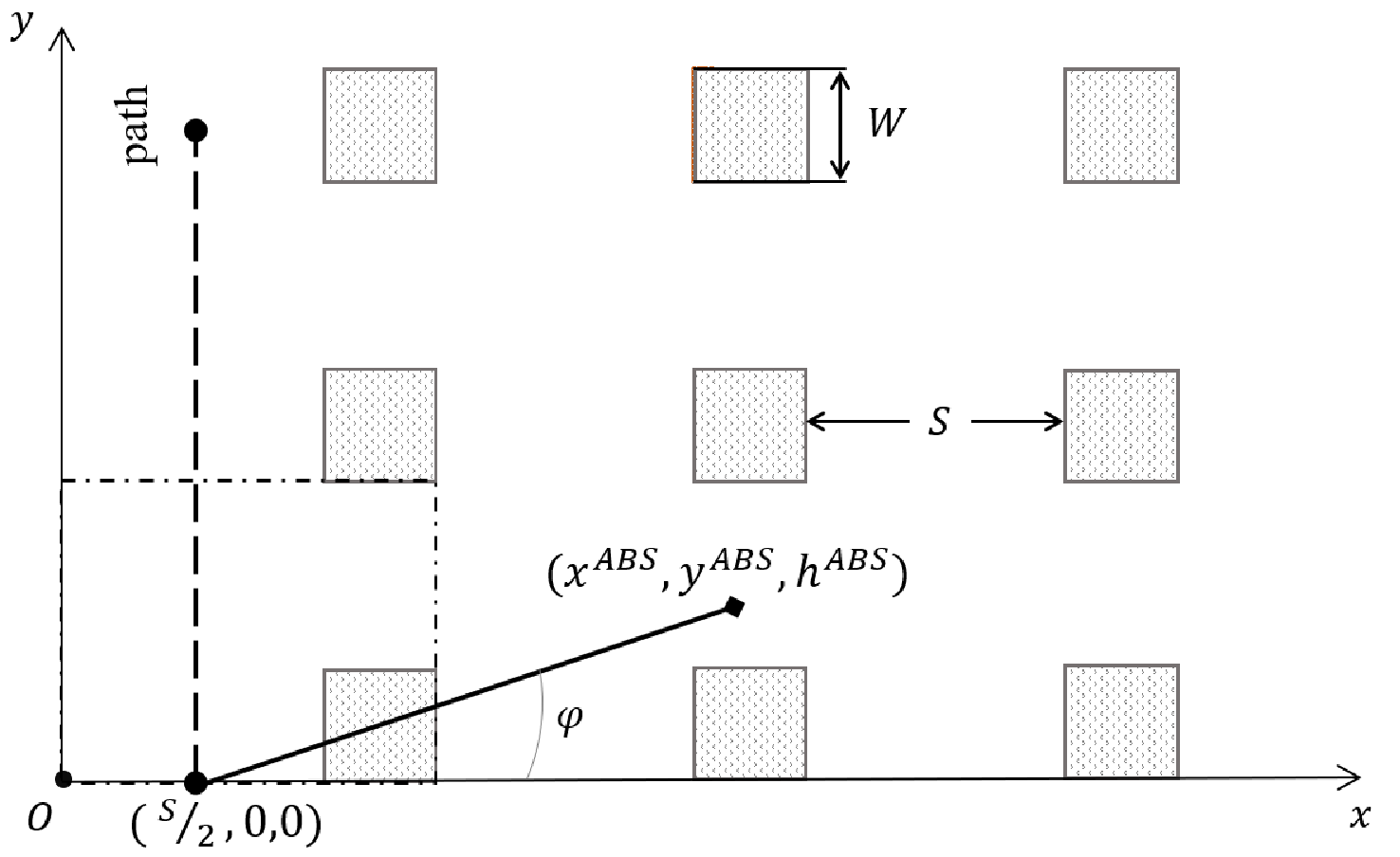}
    \caption{Manhattan-grid urban model (top view). The layout consists of multiple identical blocks (denoted with the dash-dotted lines) containing buildings and streets. Building heights are Rayleigh distributed.}
    \label{fig:layout}
    \vspace{-0.3cm}
\end{figure}

\paragraph{Manhattan grid model}
A synthetic urban layout (see Fig.~\ref{fig:layout}) is generated using environmental parameters $\alpha$, $\beta$, and $\gamma$ according to the ITU recommendations~\cite{ITU}. Each layout consists of $N$ square buildings with width $W$, arranged in a regular grid and separated by streets of width $S$. The sizes are computed as:
\begin{align}
W = 1000 \sqrt{\frac{\alpha}{\beta}}, && S = \frac{1000}{\sqrt{\beta}} - W.
\end{align}
Building heights are independently drawn from a Rayleigh distribution with scale parameter $\gamma$. The full layout contains $N = I \cdot J$ buildings, where $I$ and $J$ are the numbers of buildings along the $x$ and $y$ axes, respectively.

\paragraph{Path}
We consider a linear UE trajectory defined by two endpoints: $(\frac{S}{2}, 0)$ and $(\frac{S}{2}, K(S + W))$, where $K$ is the number of Manhattan blocks crossed. Following previous work~\cite{Saboor2023simulator,Saboor2024model}, we leverage the symmetry and homogeneity of the grid: (i) the analysis is restricted to azimuth angles $\varphi \in [0, 90]$, and (ii) the UE path is assumed linear, as more complex trajectories can be approximated by piecewise-linear segments.

\subsection{Probabilistic LOS modeling}
Creating a full 3D representation of the propagation environment is often impractical due to modeling complexity. While probabilistic LOS models for A2G links are well-established~\cite{Saboor2024model,Saboor2025pedestrian}, they typically assume independent LOS/NLOS realizations across space, neglecting spatial consistency. In reality, nearby UE locations are more likely to share the same LOS state. To address this, we define segments of the path within which the LOS/NLOS state remains constant. The segment lengths can be determined from empirical data or simple geometric models to ensure spatially consistent link behavior.

\section{LOS Model}\label{sec:los_model}

While 3D geometry-based modeling provides accurate spatial consistency of LOS obstructions, it requires a detailed 3D representation of the environment, which is often impractical. Probabilistic LOS modeling offers a more lightweight alternative by leveraging statistical distributions to determine LOS conditions along the UE trajectory.

\subsection{Path-Based LOS State Generation}

The probabilistic model proposed in Algorithm~\ref{alg:los} determines the transitions between the LOS and NLOS states for an A2G communication scenario. It operates by iterating through discrete blocks of the UE path consisting of one building and the corresponding streets (the dash-dotted line in Fig.~\ref{fig:layout}), assigning LOS and NLOS segments based on probabilistic evaluation of obstructions caused by the urban environment.

The path {of length $d_p$} is split into $K=\lfloor \frac{d_p}{S+W}\rfloor$ blocks. For each block $k$, the UE is initially positioned at $(\frac{S}{2},(k-1)(S+W),0)$. First, we compute the probability of LOS being obstructed by the closest building\footnote{As it is well demonstrated by the theoretical derivations in previous papers~\cite{Saboor2024model,Saboor2025pedestrian}, the closest buildings have the highest impact on LOS.} using \eqref{eq:PNLOS}. When the initial state is classified as NLOS, the corresponding region is assigned a length equal to the building width~$W$, and the UE position is updated accordingly (i.e., $y^{UE}=(k-1)(S+W)+W$. Next, we iteratively evaluate LOS conditions until the UE reaches the boundary of the trajectory block. At each step, the LOS probability is re-evaluated at the updated UE position. If LOS is confirmed, the length of the LOS region is sampled from a uniform distribution $\mathcal{U}(0,S)$ constrained by the street width, and the UE position is updated accordingly. Otherwise, an NLOS region is generated with a length sampled from a uniform distribution within the building width range $\mathcal{U}(0,W)$, and the UE position is advanced accordingly\footnote{Note that the model can generate several sequential segments with the same LOS/NLOS state. Although a single building shadow does not exceed $W$, several shadows can overlap, resulting in longer NLOS segments some of which can cover the whole block (and even several blocks or the whole path). The same reasoning is applied to the LOS segments}.

Upon reaching the end point of the path, the algorithm outputs the segmented LOS and NLOS regions. The resulting sequence of states can be utilized for further statistical modeling and communication performance analysis.

This probabilistic approach balances computational efficiency and accuracy, making it suitable for large-scale Monte Carlo simulations and urban mobility analysis. 

\subsection{LOS Probability Calculation}
To keep the computational complexity low, we account for the influence of only the two nearest buildings to the UE on $P_{LOS}=1-P_{NLOS}$. Note that we only consider the buildings lying on the horizontal line connecting the UE and ABS locations. Following our work in \cite{Saboor2024model}, we estimate the probability of obstruction caused by the $\kappa$-th building ($P_{NLOS}^\kappa$), assuming the UE height is $h_U = 0$ and $\kappa \in \{1, 2\}$ as
\begin{equation} 
\label{eq:PNLOS} 
P_\mathrm{NLOS}^\mathrm{\kappa}(\theta, S, W) = \frac{\gamma \cdot \sqrt{\frac{\pi}{2}}}{h_\mathrm{Bmax}^1}\cdot \left({\rm erf}\left(\frac{h_\mathrm{Bmax}^\kappa}{\sqrt{2}\gamma}\right)-{\rm erf}\left(\frac{h_\mathrm{Bmin}^\kappa}{\sqrt{2}\gamma}\right)\right)
\end{equation}
where $h_{\mathrm{Bmax}}^\kappa$ is the maximum height of the $\kappa$-th building for a given elevation angle $\theta$ that can cast its shadow across the full street width, resulting in complete NLOS condition. In contrast, $h_{\mathrm{Bmin}}^\kappa$ represents the minimum height at which the building starts obstructing any part of the observed street. So, if the $\kappa$-th building height $h_B^\kappa$ is below $h_\mathrm{Bmin}^\kappa$, it will not cast any shadow, and all UEs will remain in LOS state. For the remaining heights, we can observe partial blockage. We can compute both values using the equations below:
\begin{equation}\label{eq:heights}
\begin{aligned}
h_{\mathrm{Bmin}}^\kappa &= (\kappa - 1) \cdot (S + W) \cdot \tan(\theta), \\
h_{\mathrm{Bmax}}^\kappa &= \left[(\kappa - 1) \cdot (S + W) + S\right] \cdot \tan(\theta).
\end{aligned}
\end{equation}

The earlier expression \eqref{eq:PNLOS} assumes an azimuth angle $\varphi = 0$, which limits the model's accuracy. In reality, an ABS can communicate with the UE from various directions, resulting in changes in the effective building and the street widths. To generalize the model for any azimuth, we incorporate azimuth-dependent street and building projections:
\vspace{-0.2cm}
\begin{align}
S'(\varphi) =  S + 2S\cot(\varphi), && W'(\varphi) = \frac{W}{\cos(\varphi)}
\label{eq:StreetWidths}
\end{align}
\vspace{-0.2cm}
These effective dimensions are substituted into \eqref{eq:PNLOS} to compute azimuth-adjusted LOS probabilities.

\begin{algorithm}[H]
\caption{High-Level LOS/NLOS Generation for A2G Communication}\label{alg:los}
\begin{algorithmic}[1]
\State \textbf{Input:} Urban environment parameters, UAV position, trajectory length.
\State \textbf{Output:} Sequence of LOS/NLOS segments
\For{each block $k$ of the trajectory}
    \State Initialize UE at $(\frac{S}{2},(k-1)(S+W),0)$.
    \State Sample LOS/NLOS state using NLOS probability \eqref{eq:PNLOS} for $\kappa = 1$.
    \If{State = NLOS}
        \State Append NLOS segment of length $W$
        \State  $y^{UE} \leftarrow (k-1)(S+W)+W$
    \EndIf
    \While{the user reaches $y^{UE}=k(S+W)$}
        \State Sample LOS/NLOS state using NLOS probability \eqref{eq:PNLOS} for $\kappa = 2$.
        \If{LOS is detected}
            \State Generate LOS segment length from the uniform distribution $\ell \sim \mathcal{U}(0,S)$ 
            
        \Else
            \State Generate NLOS segment length from the uniform distribution $\ell \sim \mathcal{U}(0,W)$ 
        \EndIf
        \State Update the UE coordinates $y^{UE} \leftarrow y^{UE} + \ell$
    \EndWhile
\EndFor
\State \textbf{Return:} Final lists of LOS and NLOS segments.
\end{algorithmic}
\end{algorithm}

\section{Results}
We evaluate the proposed framework across four representative environments defined by ITU~\cite{ITU}: Suburban, Urban, Dense Urban, and High-Rise Urban. The corresponding environmental parameters, as well as other relevant simulation details, are summarized in Table~\ref{tab:setup}. These settings reflect progressively denser layouts with taller buildings.

Each simulation consists of 1000 independent realizations, where the UE follows a straight 1000-meter path sampled at 0.33-meter resolution. The ABS is randomly placed in each run, with horizontal coordinates drawn from $x, y \sim \mathcal{U}(0,1000)$~m and altitude $h^{ABS} \sim \mathcal{U}(30,300)$m. 

Communication operates at the center frequency of 2.5~GHz. The UE's receiver sensitivity is --84.7~dBm~as per the 3GPP recommendation~\cite{3gppUE}. Strict Size-Weight-and-Power (SWAP) constraints for ABSs result in moderate transmit power levels and simple antennas with low gain. In our simulations, we assume an equivalent isotropically radiated power (EIRP) of 13, 18, and 23~dBm. Path loss and shadow fading are modeled as described in Section~\ref{sec:PL_shadowing}. Outage is defined as the situation where the channels have a higher loss than $\Lambda_{outage}$ allowed by the receiver sensitivity and the transmitter EIRP: $\Lambda_{outage}=$EIRP -- UE sensitivity.

\begin{table}[]
\caption{Parameters}\label{tab:setup}
\begin{center}
\begin{tabular}
{c|c|c|c|c}
&Suburban & Urban & Dense& High-Rise\\
&&&Urban&Urban \\ \hline
$\alpha$ & 0.1 & 0.3 & 0.5& 0.5 \\ 
$\beta$ & 750 & 500 & 300& 300 \\ 
$\gamma$ & 8 & 15 & 20& 50 \\ 
$S$ &25 & 20.2 & 16.9& 16.9 \\ 
$W$ &11.5 & 24.5 & 40.8& 40.8 \\ 
\hline
No. realizations & \multicolumn{4}{c}{1000}\\
ABS coordinates & \multicolumn{4}{c}{$x$ and $y \sim \mathcal{U}(0,1000)$~m, $h^{ABS} \sim \mathcal{U}(30,300)$~m}\\
Path length $d_p$ & \multicolumn{4}{c}{1000 m}\\
Path resolution $\Delta d_p$& \multicolumn{4}{c}{0.3 m}\\
\hline
Frequency & \multicolumn{4}{c}{2.5~GHz}\\
$\Lambda_0$ & \multicolumn{4}{c}{Eq.~\eqref{eq:refPL}}\\
$\rho$ & \multicolumn{2}{c}{LOS: 0.0272}& \multicolumn{2}{c}{NLOS: 2.3197}\\
$\mu$ & \multicolumn{2}{c}{LOS: 0.7475}& \multicolumn{2}{c}{NLOS: 0.2361}\\
$d_{decorr}$ & \multicolumn{4}{c}{11 m~\cite{Bucur2019LargeScale}}\\
UE sensitivity & \multicolumn{4}{c}{-84.7~\cite{3gppUE}}\\
ABS EIRP & \multicolumn{4}{c}{13, 18, 23 dBm}\\
\hline
\end{tabular}
\label{tab1}
\end{center}
\end{table}

\subsection{Lightweight Geometry-Based Simulator}\label{sec:sim}

\begin{figure}
    \centering
    \includegraphics[width=.84\columnwidth]{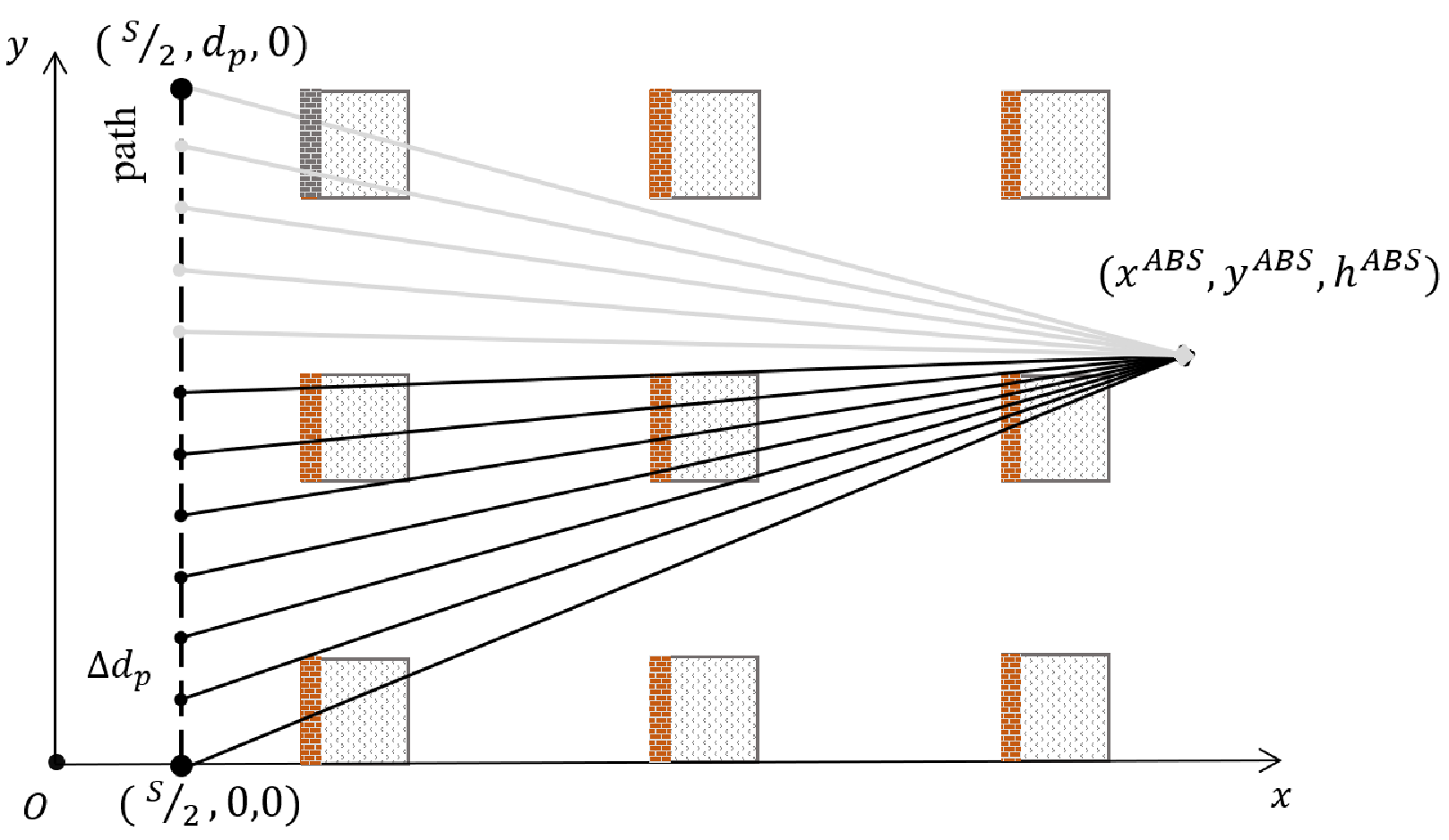}
    \caption{Simulator functionality: UE is sequentially located along the linear path with the resolution $\Delta d_p$. For each UE location, the simulator checks LOS obstructions line-of-sight by the environment. In this example, the building in the top-left corner is not tall which results in LOS.}
    \label{fig:simulator_ex}
    \vspace{-0.2cm}
\end{figure}

The model is verified against our geometry-based simulator \cite{Saboor2023simulator} which was used to validate several theoretical models in~\cite{Saboor2024model,Saboor2025pedestrian}. Although the results in~\cite{Saboor2023simulator} were generated for UEs located within a single block, the simulator has been enhanced with the functionality that allows it to take any UE location as input. Moreover, to speed up the simulations, buildings are approximated by a single wall as shown in Fig.~\ref{fig:simulator_ex}. This simplification halved the computation time while having no effect on the LOS probability calculation precision for elevation angles exceeding 20 degrees. For lower elevation angles, the error does not exceed 6\%.

Fig.~\ref{fig:simulator_ex} shows an example of the simulated environment and resulting LOS/NLOS segments (gray and black lines, respectively) of the UE path. This lightweight geometry-based method enables computationally efficient, spatially consistent LOS modeling without resorting to full-scale ray tracing.

\subsection{Spatially consistent LOS}
\begin{figure}[t!]
    \centering
    \includegraphics[width=0.9\columnwidth]{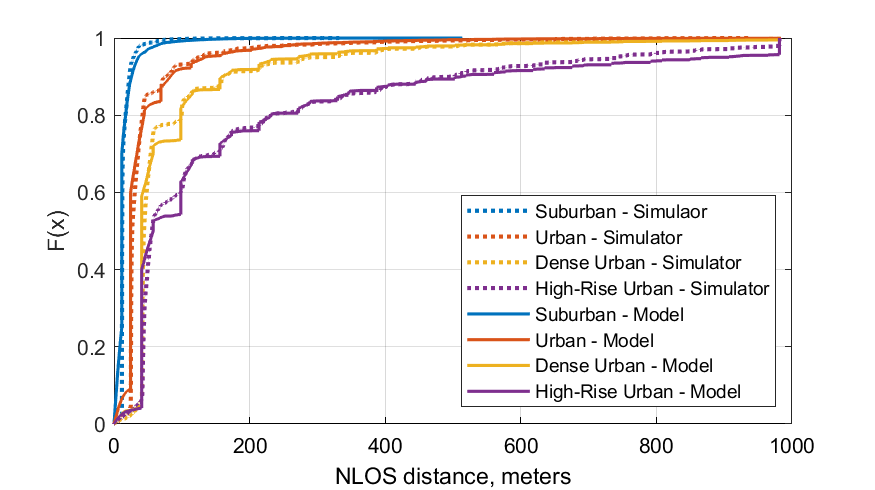}
    \caption{CDF of simulated and modeled NLOS distances across different urban environments. The step-like features correlate with the fixed building width $W$ and street width $S$ in a Manhattan layout, illustrating that sparser environments yield shorter NLOS segments.}
    \vspace{-0.2cm}
    \label{fig:nlos_cdf}
\end{figure}

Fig.~\ref{fig:nlos_cdf} compares the distribution of NLOS segment lengths across the four environments. We observe good agreement between the probabilistic model and the geometry-based simulator. Both approaches reflect the regular Manhattan grid layout, where fixed building and street widths define the obstruction intervals. The effect of the environment is also correctly reproduced: an increase in the building density and heights results in longer NLOS distances.

\subsection{Channel and Outage}
{For each point $\mathbf{p}_i$ on the path $\mathbf{P}$, our model in Section~\ref{sec:los_model} assigns a LOS state. Next, we need to generate path loss and shadow fading accordingly.}
\subsubsection{Path loss and Shadow fading models}\label{sec:PL_shadowing}
{We base our channel model on equation \eqref{eq:total_channel} where the path loss and shadow fading are modeled as in~\cite{Feng2006a2gPL}\footnote{The distances and environments investigated by Feng et al.~\cite{Feng2006a2gPL} are similar to our setup.}. The total path loss accounts for i) the reference free-space attenuation depending on the ABS height and ii) the elevation-dependent excess path loss. Together, these components reflect the link geometry in the 3D space. Similarly, shadow fading is characterized by the elevation-dependent standard deviation calculated differently for LOS and NLOS links. To ensure spatial consistency of shadow fading, we use an autoregressive (AR) process to generate shadow fading values with autocorrelation as in \eqref{eq:acf} and AR innovations following the Normal distribution with LOS/NLOS-dependent standard deviation.}

The reference distance is defined as the relative height of the ABS directly above the UE (i.e., $d_0=h^{ABS}$ in our case). Hence
\begin{equation}\label{eq:refPL}
\Lambda_0=20 \log_{10}\frac{{4\pi h^{ABS} f}}{c} \end{equation}
The excess path loss in \cite{Feng2006a2gPL} depends only on the elevation angle $\theta$ and is calculated for LOS as:
\begin{equation}
    \Lambda_{ex,LOS}(\theta) = - 20 \log_{10} \sin \theta
\end{equation}
while the NLOS excess path loss is
\begin{equation}
    \Lambda_{ex,NLOS}(\theta) = -16.16 + 12.0436 \exp(-\frac{90-\theta}{7.52}).
\end{equation}

Finally, the elevation-dependent shadow fading standard deviation for LOS and NLOS respectively is modeled as
\begin{equation}
    \sigma^\xi(\theta) = \rho (90-\theta)^\mu,
\end{equation}
where the parameters $\rho$ and $\mu$ can be found in Table~\ref{tab:setup} for the LOS and NLOS cases. The decorrelation distance is found to be varying between 9.5 and 12.9~m~\cite{Bucur2019LargeScale}, we use 11~m.

\subsubsection{Outage analysis}
\begin{figure}[t!]
    \centering
    \includegraphics[width=.9\columnwidth]{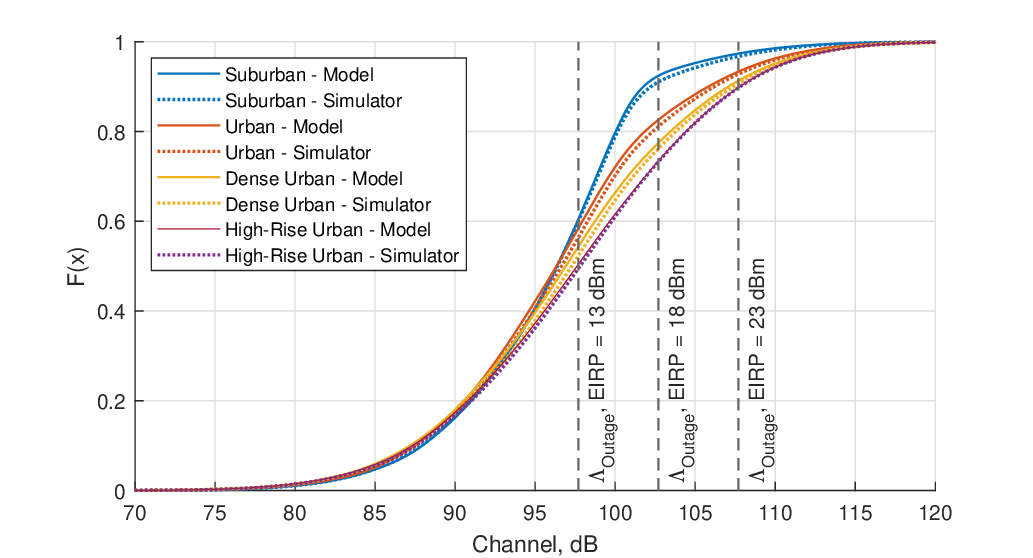}
    \caption{CDF of channel composed on path loss and shadowing}
    \label{fig:PL_shadowing}
    \vspace{-0.2cm}
\end{figure}

Fig.~\ref{fig:PL_shadowing} shows channel attenuation obtained with our model and simulator for different environments. We observe that for EIRP = 23 dBm, outage probability does not exceed 10.3\% (in High-Rise Urban layout) and go as low as 3\% in Suburban environment. However, for ABS EIRP = 13 dBm, we observe 51\% and 40\% outage probability in the same setups.

\begin{figure}[h!]
    \centering
    \includegraphics[width=0.45\textwidth]{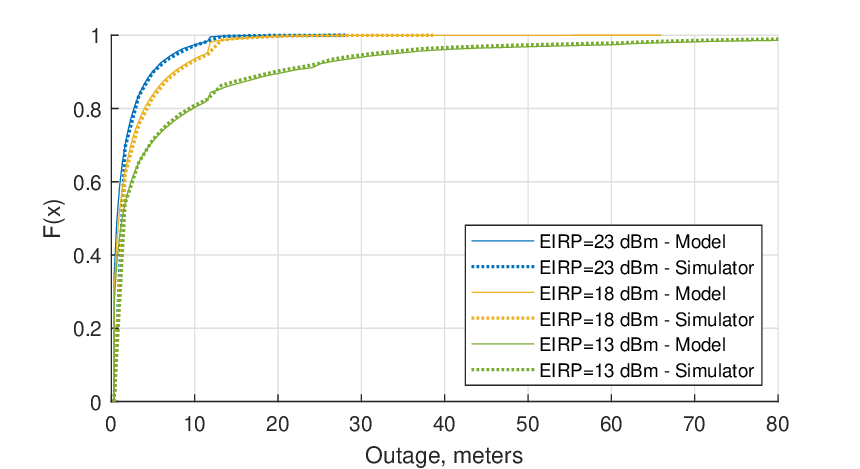}
    \includegraphics[width=0.45\textwidth]{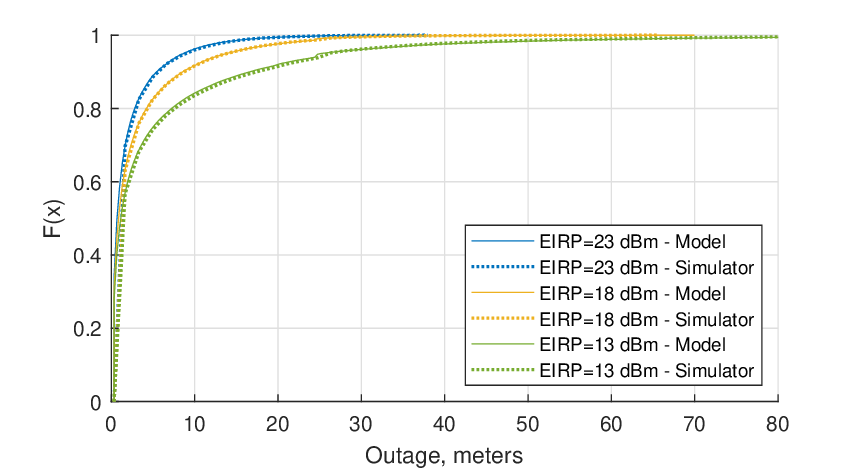}
    \includegraphics[width=0.45\textwidth]{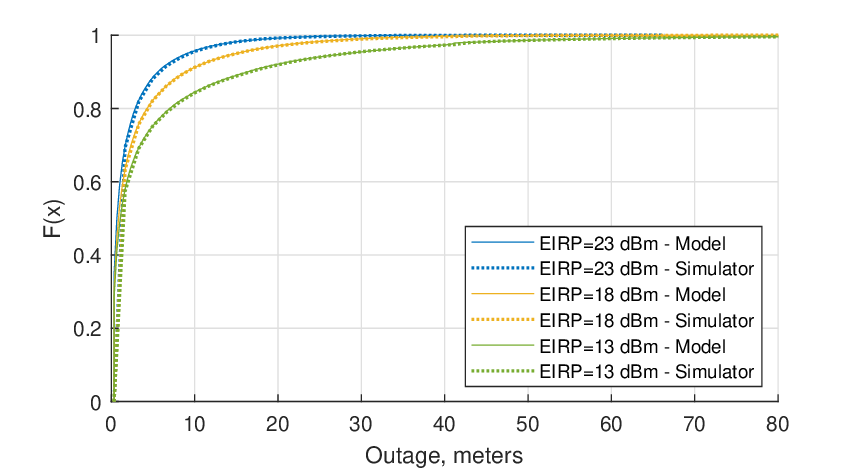}
    \includegraphics[width=0.45\textwidth]{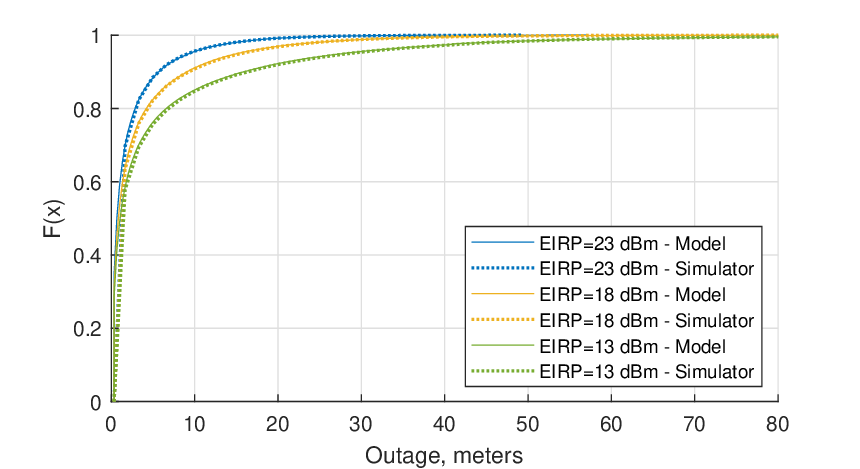}
    \caption{CDF of Outage distances for i) Suburban ii) Urban iii) Dense Urban iv) High-Rise Urban}
    \label{fig:outage_distance}
    \vspace{-0.5cm}
\end{figure}

Similar to Average Fade Duration (AFD) used to characterize small-scale fading, we define outage distance {as segments of the UE path where the received signal power drops below the receiver sensitivity. The outage distance statistics are} shown in Fig.~\ref{fig:outage_distance}. In the suburban environment, we observe that the outage distances for EIRP=23~dBm rarely exceed the building width. The outage distances grow once EIRP is decreased or a more dense environment is considered, however, the distances do not exceed 40 meters with 95\% probability. Interestingly, the effect of building width is obvious only in suburban areas while only very subtle bumps at $W$ and $S$ can be observed in other environments for EIRP=13~dBm. Moreover, there is no difference between Dense and High-Rise Urban environments in terms of outage distances indicating that the density of buildings is more influential than the building heights. 

\section{Conclusions}
This paper presented a spatially consistent A2G channel modeling framework using probabilistic LOS/NLOS segmentation to parameterize the deterministic path loss and stochastic shadow fading model. The proposed approach enables the generation of correlated large-scale fading across space without relying on detailed 3D city models, offering a low-complexity and versatile tool for the performance evaluation of drone-assisted wireless systems. Our results demonstrate that the model effectively captures realistic LOS/NLOS patterns and produces channel characteristics that are consistent with those observed in geometry-based simulations.

\bibliographystyle{IEEEtran}
\bibliography{main}

\end{document}